\documentclass[10pt,american,english,aps,showkeys,showpacs,nofootinbib]{revtex4-1}
\usepackage[LGR,T1]{fontenc}
\usepackage{textcomp}
\usepackage[utf8]{inputenc}
\usepackage{babel}
\usepackage{amsmath}
\usepackage{amssymb}
\usepackage{stackrel}
\usepackage{graphicx}
\usepackage{geometry}
\usepackage[bookmarks=false,
 breaklinks=false,pdfborder={0 0 1},backref=section,colorlinks=false]
 {hyperref}

\makeatletter

\DeclareRobustCommand{\greektext}{%
  \fontencoding{LGR}\selectfont\def\encodingdefault{LGR}}
\DeclareRobustCommand{\textgreek}[1]{\leavevmode{\greektext #1}}

\usepackage{babel}
\usepackage{todonotes}
\usepackage[bottom]{footmisc}
\usepackage{graphicx}
\usepackage{dcolumn}
\usepackage{bm}
\usepackage{todonotes}
\ifdefined\showcaptionsetup
 \PassOptionsToPackage{caption=false}{subfig}
\fi
\usepackage{subfig}
\makeatother

\begin{document}
\title{Analysis of Spin-1/2 Particle Scattering in a Spinning Cosmic String
Spacetime with Torsion, Curvature, and a Coulomb Potential}
\author{Abdelmalek Boumali}
\email{boumali.abdelmalek@gmail.com}

\affiliation{Laboratory of theoretical and applied Physics~\\
 Echahid Cheikh Larbi Tebessi University, Algeria}
\date{\today}
\selectlanguage{american}%
\begin{abstract}
This paper investigates the scattering states of spin-(1/2) particles
in the spacetime of a spinning cosmic string with spacelike disclination
and dislocation, with and without a Coulomb interaction. Working within
the tetrad formalism, we solve the Dirac equation for several configurations
of angular momentum density $J_{t}$ and the torsion parameter $J_{z}$
that are relevant from a physical perspective. These configurations
include balanced torsion ($J_{t}=J_{z}$), pure spinning strings ($J_{z}=0$),
pure screw dislocations ($J_{t}=0$) and the general case. In all
cases, the geometry modifies an effective azimuthal quantum number,
and for strong rotation it introduces a geometric radial cutoff $\rho_{c}$
that acts as a hard wall. These factors lead to closed-form expressions
for the radial wave functions, phase shifts and differential cross
sections, which are expressed in terms of confluent hypergeometric
and Bessel functions. We demonstrate that conical curvature, rotation
and torsion generate Aharonov--Bohm--like contributions, as well
as energy- and momentum-dependent asymmetries in Dirac--Coulomb scattering.
This results in topology-renormalised Mott/Rutherford patterns. In
the Coulomb-free limit, scattering becomes purely geometric yet still
exhibits characteristic forward enhancement, which is governed by
defect parameters and the cutoff. We briefly discuss possible realisations
in Dirac materials, such as strained or defective graphene, where
lattice disclinations and dislocations mimic the cosmic-string geometry.
\end{abstract}
\maketitle
\selectlanguage{english}%

\section{Introduction}

Topological defects, such as cosmic strings, are hypothetical remnants
of symmetry-breaking phase transitions in the early universe as predicted
by grand unified theories. They induce conical curvature (disclination),
which is characterised by an angular deficit $\alpha$, and when they
are endowed with intrinsic spin or lattice-like distortions, they
may also carry torsion/dislocation, which is encoded by rotation and
screw-dislocation parameters ($J_{t}$ and $J_{z}$). In cylindrical
coordinates, these geometric features enter Dirac theory via the tetrad
and spin connection. In strongly rotating regimes, they can generate
a geometry-induced lower radial bound $\rho_{c}$ that excludes non-causal
regions and acts as a hard-wall boundary. Consequently, the partial-wave
indices, phase shifts and scattering observables for spin-(1/2) particles
are all modified by the underlying spacetime structure.

Most previous investigations of fermions in cosmic-string backgrounds
have focused on bound states and spectral properties, demonstrating
the effects of the angular deficit and torsion on Dirac energy levels,
effective centrifugal barriers, and spin--orbit couplings. In contrast,
scattering states which directly encode observable cross sections
and potential signatures of strings in astrophysical or analogue condensed-matter
settings have received considerably less attention, especially when
combined with long-range Coulomb fields. However, in realistic scenarios,
charged sources or electromagnetic fields are expected to coexist
with string-like defects and their interaction with curvature, rotation
and torsion can generate distinctive diffraction patterns and phase-shift
structures that are not captured by bound-state analyses alone \cite{KatanaevVolovich1992,FischerVisser2002,Carneiro2023,Kibble1976,Vilenkin1981,VilenkinShellard1994,ClarkeEllisVickers1990,MarquesBezerra2002,Chen2020,BoumaliMessai2014,BoumaliMessai2017,MessaiBoumali2015,MessaiBoumali2025,Rouabhia2023}.

The main goal of this paper is to provide a unified, analytical description
of Dirac scattering in a spinning cosmic-string spacetime with spacelike
disclination and dislocation, with and without a static Coulomb potential.
We demonstrate how the geometry reshapes scattering through two key
factors \cite{PuntigamSoleng1997,Ozdemir2005,Jusufi2016}:

(i) an effective azimuthal quantum number $\kappa_{eff}(\alpha,J_{t},J_{z};E,k)$
that incorporates the conical angle, rotation, torsion and the energy--momentum
of the particle; and

(ii) a geometric cutoff $\rho_{c}$, which is activated in strongly
rotating regimes and introduces additional boundary-induced phases.
Together, these quantities control the partial-wave phase shifts and
thus the differential and total cross sections.

To clarify the function of each geometric parameter, we analyse several
physically relevant configurations: the pure cosmic string ($J_{t}=J_{z}=0$),
the balanced torsion case ($J_{t}=J_{z}$), the pure spinning string
($J_{z}=0$), the pure screw dislocation ($J_{t}=0$), and the fully
general case with both $J_{t}$ and $J_{z}$ non-zero. For each configuration,
we derive the radial Dirac equations using the tetrad formalism and
solve them in closed form: confluent hypergeometric functions describe
the Coulomb problem, while Bessel functions describe purely geometric
scattering. This enables us to derive explicit expressions for the
phase shifts, scattering amplitudes and cross sections and to determine
how conical curvature, rotation, torsion and the cutoff $\rho_{c}$
generate Aharonov--Bohm--like contributions and energy- or momentum-dependent
asymmetries. Additionally, we briefly discuss possible realisations
in Dirac materials, such as strained or defective graphene and Dirac/Weyl
semimetals, where lattice disclinations and dislocations can mimic
the cosmic-string geometry and render the predicted scattering effects
accessible to experimentation.

For clarity, the paper is organised as follows. Section II introduces
the spinning cosmic-string metric with spacelike disclination and
dislocation, reviews the tetrad formalism and derives the Dirac equation
in cylindrical coordinates. This section highlights the emergence
of $\kappa_{eff}$ and the geometric cutoff $\rho_{c}$. Section III
treats Dirac scattering in the presence of a Coulomb potential. It
analyses the pure string, balanced torsion, pure spinning, screw-dislocation
and general cases in detail, deriving the corresponding radial solutions
and phase shifts. Section IV focuses on geometric scattering in the
absence of the Coulomb interaction, emphasising the purely topological
contributions to the scattering amplitude. Section V presents and
compares the angular distributions and total cross sections for all
configurations, illustrating how curvature, rotation, torsion and
the cutoff modify the Mott/Rutherford patterns and the Coulomb-free
diffraction structures. Section VI discusses potential experimental
implementations in graphene and other Dirac materials, where analogue
cosmic-string geometries may be engineered. Finally, Section VII summarises
our main results and outlines possible extensions, including additional
interactions and dynamical string backgrounds.
\selectlanguage{american}%

\section{DIRAC equation IN THE COSMIC STRING BACKGROUND}

\selectlanguage{english}%
In this section, we address the solution of the Dirac oscillator in
the presence of a cosmic string background characterized by the spacetime
signature $(-+++)$. The geometry induced by the cosmic string is
described using cylindrical coordinates $(t,\rho,\phi,z)$, and the
corresponding line element for a straight, rotating cosmic string
endowed with torsion is given by \cite{PuntigamSoleng1997,Ozdemir2005,Jusufi2016}:
\begin{equation}
ds^{2}=-\left(dt+4GJ^{t}d\varphi\right)^{2}+d\rho^{2}+\alpha^{2}\rho^{2}d\varphi^{2}+\left(dz+4GJ^{z}d\varphi\right)^{2},\label{1}
\end{equation}
where the parameter $\alpha$, satisfying $0<\alpha<1$, represents
the angular deficit associated with the conical geometry. The quantity
$J^{t}$ denotes the linear density of angular momentum, responsible
for frame-dragging effects, while $J^{z}$ characterizes the screw-dislocation
parameter linked to torsion. 

The coordinate ranges are defined as $-\infty<t,z<\infty$, $0\leq\rho<\infty$,
and $0\leq\phi\leq2\pi$. The deficit parameter $\alpha$ is associated
with the conical structure of the spacetime and satisfies the relation
$\alpha=1-4\mu$, where \$\textbackslash mu\$ is the linear mass
density of the cosmic string expressed in natural units.

\selectlanguage{american}%
The metric tensor $g_{\mu\nu}$ in matrix form is:
\begin{equation}
g_{\mu\nu}=\begin{pmatrix}-1 & 0 & -4GJ^{t} & 0\\
0 & 1 & 0 & 0\\
-4GJ^{t} & 0 & \alpha^{2}\rho^{2}-16G^{2}[(J^{t})^{2}-(J^{z})^{2}] & 4GJ^{z}\\
0 & 0 & 4GJ^{z} & 1
\end{pmatrix}\label{2}
\end{equation}
We note here that for the metric in Eq. (\ref{1}), the azimuthal
component is $g_{\phi\phi}=\alpha^{2}\rho^{2}-16G^{2}[(J_{t})^{2}-(J_{z})^{2}]$.
Physical admissibility requires $g_{\phi\phi}>0$ to ensure a positive
definite metric signature and avoid closed timelike curves or unphysical
regions\cite{Vitoria2025_IJMPA_2450176,Bakke2020_IJMPA_2050129}.
Hence, when $(J_{t})^{2}>(J_{z})^{2}$, we restrict the configuration
space to 
\begin{equation}
\rho>\rho_{c}=\frac{4G}{\alpha}\sqrt{(J_{t})^{2}-(J_{z})^{2}}.\label{2-1}
\end{equation}
For $J_{z}=0$, this reduces to $\rho>4G|J_{t}|/\alpha$; for $J_{t}=0$,
the inequality is automatically satisfied for all $\rho>0$. 

This radial cutoff $\rho_{c}$ has important implications for the
quantum system. All normalization integrals for the wavefunctions
are thus evaluated over $\rho\in(\rho_{c},\infty)$ (or $(0,\infty)$
when $\rho_{c}=0$), ensuring that the probability density is confined
to the physically admissible region. Boundary conditions at $\rho=\rho_{c}$
are imposed such that the wavefunction vanishes or satisfies the conditions
that prevent leakage into the forbidden zone, analogous to hard-wall
potentials in defect spacetimes. This restriction modifies the effective
centrifugal barrier in the radial equation, potentially shifting the
energy levels and altering the density of states. In particular, for
large torsion parameters, $\rho_{c}$ introduces a minimal radius
that lifts low-angular-momentum degeneracies and affects the ground-state
energy, providing a geometric regularization akin to those in rotating
frames \cite{Vitoria2025_IJMPA_2450176,Bakke2020_IJMPA_2050129}.

\selectlanguage{english}%
Now, the physical properties of this spacetime are determined by the
value of $j^{2}=(J^{t})^{2}-(J^{z})^{2}$ \cite{Ozdemir2005}: 
\begin{itemize}
\item Case (1): $j^{2}=0$ (i.e., $|J^{z}|=|J^{t}|$)
\begin{equation}
ds^{2}=-\left(dt+4GJ\,d\varphi\right)^{2}+d\rho^{2}+\alpha^{2}\rho^{2}d\phi^{2}+\left(dz+4GJ\,d\varphi\right)^{2}\label{3}
\end{equation}
with
\begin{equation}
g_{\mu\nu}=\begin{pmatrix}-1 & 0 & -4GJ^{t} & 0\\
0 & 1 & 0 & 0\\
-4GJ^{t} & 0 & \alpha^{2}\rho^{2} & 4GJ^{z}\\
0 & 0 & 4GJ^{z} & 1
\end{pmatrix}\label{4}
\end{equation}
This describes a string interacting with a circularly polarized plane-fronted
gravitational wave.
\item Case (2): $|J^{z}|=0$
\begin{equation}
ds^{2}=-\left(dt+4GJ^{t}d\varphi\right)^{2}+d\rho^{2}+\alpha^{2}\rho^{2}d\varphi^{2}+dz^{2}\label{eq:5}
\end{equation}
with
\begin{equation}
g_{\mu\nu}=\begin{pmatrix}-1 & 0 & -4GJ^{t} & 0\\
0 & 1 & 0 & 0\\
-4GJ^{t} & 0 & \alpha^{2}\rho^{2}-16G^{2}(J^{t})^{2} & 0\\
0 & 0 & 0 & 1
\end{pmatrix}\label{6}
\end{equation}
This corresponds to a spinning cosmic string without any spatial dislocation.
\item Case (3): $|J^{t}|=0$
\begin{equation}
ds^{2}=-dt^{2}+dr^{2}+\alpha^{2}\rho^{2}d\varphi^{2}+\left(dz+4GJ^{z}d\varphi\right)^{2}\label{7}
\end{equation}
with
\begin{equation}
g_{\mu\nu}=\begin{pmatrix}-1 & 0 & 0 & 0\\
0 & 1 & 0 & 0\\
0 & 0 & \alpha^{2}\rho^{2}+16G^{2}(J^{z})^{2} & 4GJ^{z}\\
0 & 0 & 4GJ^{z} & 1
\end{pmatrix}\label{8}
\end{equation}
In this scenario, the spacetime describes screw dislocations, which
can be interpreted as a combination of a screw dislocation (with $2GJ^{z}/\pi$
analogous to a Burgers vector) and a disclination
\end{itemize}
The governing equation for the spinor field in this curved background
is the Dirac equation \cite{ObukhovHehl2004,Pollock2010,Fulling1989,Nakahara2003,ParkerToms2009,PuntigamSoleng1997}:
\begin{equation}
\left[i\gamma^{\mu}(x)\left(\partial_{\mu}-\Gamma_{\mu}(x)\right)-m\right]\Psi(t,x)=0,\label{9}
\end{equation}
which differs from its flat spacetime counterpart due to the presence
of the additional term $\gamma^{\mu}(x)\Gamma_{\mu}(x)$, accounting
for the geometric effects introduced by the conical defect. 

The generalized gamma matrices $\gamma^{\mu}(x)$ satisfy the Clifford
algebra $\{\gamma^{\mu},\gamma^{\nu}\}=2g^{\mu\nu}$, and are expressed
in terms of the standard Dirac matrices $\gamma^{a}$ in Minkowski
space via the tetrad fields as:
\begin{equation}
\gamma^{\mu}(x)=e_{a}^{\mu}(x)\gamma^{a}.\label{10}
\end{equation}
The tetrads $e_{\mu}^{a}(x)$ fulfill the orthonormality condition:
\begin{equation}
e_{\mu}^{a}(x)e_{\nu}^{b}(x)\eta_{ab}=g_{\mu\nu},\label{11}
\end{equation}
where the indices $\mu,\nu=0,1,2,3$ refer to curved spacetime coordinates,
and $a,b=0,1,2,3$ denote flat spacetime (tetrad) indices. 

The spin connection $\Gamma_{\mu}(x)$ is obtained via:
\begin{equation}
\Gamma_{\mu}(x)=\frac{1}{8}\omega_{\mu ab}(x)[\gamma^{a},\gamma^{b}],\label{12}
\end{equation}
with the spin connection one-forms $\omega_{\mu ab}$ defined as:
\begin{equation}
\omega_{\mu ab}=e_{a\nu}(\partial_{\mu}e_{b}^{\nu}+\Gamma_{\mu\lambda}^{\nu}e_{b}^{\lambda}).\label{13}
\end{equation}
In what follows, we will treat the three dimensional Dirac oscilltor
for each case mentioned above.

\subsection{First case: \foreignlanguage{american}{Dirac equation in a Spinning
Cosmic String Background} with Equal Angular Momentum and Torsion
($j^{2}=0$)}

We consider the Dirac in the curved background of a spinning cosmic
string with equal temporal and spatial torsion components, such that
$J^{t}=J^{z}=J$, leading to a simplified torsional configuration
$j^{2}=0$. The spacetime geometry is encoded in the tetrad fields:
\begin{equation}
e_{\mu}^{a}(x)=\begin{pmatrix}1 & 0 & 4GJ & 0\\
0 & \cos\phi & -\alpha\rho\sin\phi & 0\\
0 & \sin\phi & \alpha\rho\cos\phi & 0\\
0 & 0 & 4GJ & 1
\end{pmatrix},\quad e_{a}^{\mu}=\begin{pmatrix}1 & 0 & -\frac{4GJ}{\alpha\rho} & 0\\
0 & \cos\phi & \frac{\sin\phi}{\alpha\rho} & 0\\
0 & -\sin\phi & \frac{\cos\phi}{\alpha\rho} & 0\\
0 & 0 & -\frac{4GJ}{\alpha\rho} & 1
\end{pmatrix}.\label{14}
\end{equation}
From these, we obtain the position-dependent gamma matrices in the
curved spacetime:
\begin{equation}
\gamma^{t}=\gamma^{0}-\frac{4GJ}{\alpha\rho}\gamma^{2},\quad\gamma^{\rho}=\cos\phi\,\gamma^{1}+\sin\phi\,\gamma^{2},\quad\gamma^{\phi}=\frac{-\sin\phi\,\gamma^{1}+\cos\phi\,\gamma^{2}}{\alpha\rho},\quad\gamma^{z}=\gamma^{3}-\frac{4GJ}{\alpha\rho}\gamma^{2}.\label{15}
\end{equation}

\subsection{Second case: \foreignlanguage{american}{Dirac equation in a Purely
Spinning Cosmic String Background ($J^{z}=0$)}}

\selectlanguage{american}%
In the second configuration, we consider a purely spinning cosmic
string background, where the temporal component of torsion is retained
while the spatial component vanishes, i.e., $J^{t}\neq0$, $J^{z}=0$.
The tetrad fields reduce to:
\begin{equation}
e_{\mu}^{a}(x)=\begin{pmatrix}1 & 0 & 4GJ^{t} & 0\\
0 & \cos\phi & -\alpha\rho\sin\phi & 0\\
0 & \sin\phi & \alpha\rho\cos\phi & 0\\
0 & 0 & 0 & 1
\end{pmatrix},\quad e_{a}^{\mu}(x)=\begin{pmatrix}1 & 0 & -\frac{4GJ^{t}}{\alpha\rho} & 0\\
0 & \cos\phi & \frac{\sin\phi}{\alpha\rho} & 0\\
0 & -\sin\phi & \frac{\cos\phi}{\alpha\rho} & 0\\
0 & 0 & 0 & 1
\end{pmatrix}.\label{16}
\end{equation}
The corresponding gamma matrices become:
\begin{equation}
\gamma^{t}=\gamma^{0}-\frac{4GJ^{t}}{\alpha\rho}\gamma^{2},\quad\gamma^{\rho}=\cos\phi\gamma^{1}+\sin\phi\gamma^{2},\quad\gamma^{\phi}=\frac{-\sin\phi\gamma^{1}+\cos\phi\gamma^{2}}{\alpha\rho},\quad\gamma^{z}=\gamma^{3}.\label{17}
\end{equation}

\selectlanguage{english}%

\subsection{Third case\foreignlanguage{american}{: Dirac equation in a Cosmic
String Background with Screw Dislocations ($J^{t}=0$)}}

\selectlanguage{american}%
In the third case, the background spacetime consists of a cosmic string
with screw dislocation, where $J^{t}=0$ and $J^{z}\neq0$. This reflects
purely spatial torsion along the $z$-axis. The tetrads simplify accordingly:
\begin{equation}
e_{\mu}^{a}(x)=\begin{pmatrix}1 & 0 & 0 & 0\\
0 & \cos\phi & -\alpha\rho\sin\phi & 0\\
0 & \sin\phi & \alpha\rho\cos\phi & 0\\
0 & 0 & 4GJ^{z} & 1
\end{pmatrix},\quad e_{a}^{\mu}(x)=\begin{pmatrix}1 & 0 & 0 & 0\\
0 & \cos\phi & \frac{\sin\phi}{\alpha\rho} & 0\\
0 & -\sin\phi & \frac{\cos\phi}{\alpha\rho} & 0\\
0 & 0 & -\frac{4GJ^{z}}{\alpha\rho} & 1
\end{pmatrix}.\label{18}
\end{equation}
The corresponding gamma matrices read:
\begin{equation}
\gamma^{t}=\gamma^{0},\quad\gamma^{\rho}=\cos\phi\gamma^{1}+\sin\phi\gamma^{2},\quad\gamma^{\phi}=\frac{-\sin\phi\gamma^{1}+\cos\phi\gamma^{2}}{\alpha\rho},\quad\gamma^{z}=\gamma^{3}-\frac{4GJ^{z}}{\alpha\rho}\gamma^{2}.\label{19}
\end{equation}

\subsection{General case\foreignlanguage{english}{: Dirac equation in a Torsion
and Curvature-Modified Spacetime}}

\selectlanguage{english}%
We consider the dynamics of the Dirac  in a nontrivial spacetime background
incorporating both curvature and torsion. The metric is specified
in matrix form as
\begin{equation}
g_{\mu\nu}=\begin{pmatrix}-1 & 0 & -4GJ^{t} & 0\\
0 & 1 & 0 & 0\\
-4GJ^{t} & 0 & \alpha^{2}\rho^{2}-16G^{2}\left[(J^{t})^{2}-(J^{z})^{2}\right] & 4GJ^{z}\\
0 & 0 & 4GJ^{z} & 1
\end{pmatrix},\label{20}
\end{equation}
This metric includes both off-diagonal and curvature-corrected diagonal
components, and is consistent with cosmic string-like sources with
intrinsic spin and dislocation.

To construct a spinor formalism in this background, we seek a tetrad
field $e_{\mu}^{a}(x)$ satisfying $g_{\mu\nu}=e_{\mu}^{a}e_{\nu}^{b}\eta_{ab}$,
where $\eta_{ab}=\mathrm{diag}(-1,1,1,1)$ is the Minkowski metric
in the local Lorentz frame. 

A compatible choice is
\begin{equation}
e_{\mu}^{a}(x)=\begin{pmatrix}1 & 0 & 4GJ^{t} & 0\\
0 & 1 & 0 & 0\\
0 & 0 & \alpha\rho & 0\\
0 & 0 & 4GJ^{z} & 1
\end{pmatrix},\quad e_{a}^{\mu}(x)=\begin{pmatrix}1 & 0 & -\dfrac{4GJ^{t}}{\alpha\rho} & 0\\
0 & 1 & 0 & 0\\
0 & 0 & \dfrac{1}{\alpha\rho} & 0\\
0 & 0 & -\dfrac{4GJ^{z}}{\alpha\rho} & 1
\end{pmatrix}.\label{21}
\end{equation}
From this tetrad, the curved-space Dirac matrices are constructed
via $\gamma^{\mu}(x)=e_{a}^{\mu}(x)\gamma^{a}$, yielding:
\begin{equation}
\gamma^{t}=\gamma^{0}-\frac{4GJ^{t}}{\alpha\rho}\gamma^{2},\quad\gamma^{\rho}=\gamma^{1},\quad\gamma^{\phi}=\frac{1}{\alpha\rho}\gamma^{2},\quad\gamma^{z}=\gamma^{3}-\frac{4GJ^{z}}{\alpha\rho}\gamma^{2}.\label{22}
\end{equation}

\selectlanguage{american}%

\section{Scattering States Under Coulomb Potential}

\selectlanguage{english}%
In this section, we formulate the Dirac equation for a massive spin-1/2
particle propagating in the curved background of a spinning cosmic
string spacetime with Torsion, curvature, and a coulomb potential.
The geometry induced by the cosmic string is described using cylindrical
coordinates $(t,\rho,\phi,z)$, and the corresponding line element
for a straight, rotating cosmic string endowed with torsion is given
by \cite{PuntigamSoleng1997,Ozdemir2005,Jusufi2016}:
\selectlanguage{american}%

\subsection{Pure cosmic string $J_{t}=J_{z}=0$}

We work in natural units $\hbar=c=1$. The metric of a straight cosmic
string with deficit ($\alpha\in(0,1${]} is \cite{Carvalho2011}
\begin{equation}
ds^{2}=dt^{2}-d\rho^{2}-\alpha^{2}\rho^{2}d\phi^{2}-dz^{2},\label{23}
\end{equation}
and the (tetrad) choice gives 
\begin{equation}
\gamma^{t}=\gamma^{0},\ \gamma^{\rho}=\gamma^{1},\ \gamma^{\phi}=\gamma^{2}/(\alpha\rho),\ \gamma^{z}=\gamma^{3}\label{24}
\end{equation}
The only nonzero spin connection component is
\begin{equation}
\Gamma_{\phi}=\frac{1-\alpha}{2}\Sigma^{12},\qquad\Sigma^{ab}=\frac{1}{4}[\gamma^{a},\gamma^{b}],\label{25}
\end{equation}
which produces the well-known shift in the orbital barrier (AB-like).

Add a static Coulomb vector potential
\begin{equation}
A_{\mu}=(A_{t},0,0,0)\label{26}
\end{equation}
 with
\begin{equation}
A_{t}=-V(\rho)=-\frac{\eta}{\rho},\qquad(\eta=Z\alpha_{f}\ \text{in natural units, sign chosen for attraction}).\label{27}
\end{equation}
The covariant Dirac equation
\[
\left[i\gamma^{\mu}(x)\left(\partial_{\mu}-ieA_{\mu}\right)-i\gamma^{\mu}(x)\Gamma_{\mu}(x)-m\right]\Psi=0
\]
 With cylindrical symmetry,we take the separated ansatz
\[
\Psi(t,\rho,\phi)=e^{-iEt}e^{il\phi}e^{ik_{z}z}\begin{pmatrix}F(\rho)\ G(\rho)\end{pmatrix}\otimes\chi,
\]
with constant two-spinor $\chi$ (fix a definite spin projection).

The effect of the string is to replace ($l\to l/\alpha$) in the centrifugal
piece and add the spin connection shift ($\propto(1-\alpha)\Sigma^{12}$).
These combine into an $\alpha$-dependent Dirac index (generalized
$\kappa$)
\begin{equation}
\kappa(\alpha)=\pm\frac{j+\tfrac{1}{2}}{\alpha},\qquad j=|l|-\tfrac{1}{2},|l|+\tfrac{1}{2},\ldots\label{28}
\end{equation}
(upper/lower sign for the two spinor couplings).

With 
\begin{equation}
V(\rho)=\eta/\rho\label{29}
\end{equation}
the radial Dirac system becomes
\begin{equation}
\frac{dF}{d\rho}+\frac{\kappa(\alpha)}{\rho}F=\big(E+m-V(\rho)\big),G\label{30}
\end{equation}
\begin{equation}
\frac{dG}{d\rho}-\frac{\kappa(\alpha)}{\rho}G=\big(E-m+V(\rho)\big)F\label{31}
\end{equation}
Eliminate $G$ to get a Schrödinger-like equation for $F$ (similarly
for $G$):
\begin{equation}
\left[-\frac{d^{2}}{d\rho^{2}}+\frac{\kappa(\alpha)\big(\kappa(\alpha)-1\big)}{\rho^{2}}-\frac{2\eta E}{\rho}+\frac{2\eta\kappa(\alpha)}{\rho^{2}}\right]F(\rho)=k^{2}F(\rho)\label{32}
\end{equation}
with the continuum wavenumber ($k\equiv\sqrt{E^{2}-m^{2}}$) (scattering
regime $E>m$). The ($\rho^{-2}$) term shows how $\alpha$ and $\eta$
mix in the effective centrifugal barrier.

Introduce $z=2k\rho$ and the standard Coulomb ansatz
\begin{equation}
F(\rho)=\rho^{s}e^{-ik\rho}w(\rho),\qquad s=\sqrt{\kappa(\alpha){}^{2}-2\eta\kappa\alpha+(\eta E/k)^{2}},\label{33}
\end{equation}
which yields Kummer’s equation
\begin{equation}
zw''+(2s+1-z)w'+\left(s+i\frac{\eta E}{k}\right)w=0.\label{34}
\end{equation}
The solution regular at the origin:
\begin{equation}
F(\rho)=N(2k\rho)^{s}e^{-ik\rho}{}_{1}F_{1}\left(s+i\frac{\eta E}{k},2s+1,2k\rho\right)\label{35}
\end{equation}
with $s\in\mathbb{R}$ if $\kappa(\alpha)^{2}>(\eta E/k)^{2}$ (avoids
supercritical complications). The companion $G(\rho)$ follows from
the first-order system.

\paragraph{Asymptotics and phase shifts}

Using the large-$z$ form of $_{1}F_{1}$, the spinor behaves asymptotically
as outgoing/incoming spherical waves with a long-range Coulomb logarithm:
\begin{equation}
F(\rho)\sim\sin\left(k\rho-\frac{l\pi}{2}+\delta_{l}(\alpha)\right),\qquad\delta_{l}(\alpha)=\frac{\pi}{2}\big(|\kappa(\alpha)|-s\big)+\arg\Gamma(2s+1)-\arg\Gamma\left(s+i\frac{\eta E}{k}\right)+\frac{\eta E}{k}\ln(2k\rho).\label{36}
\end{equation}
The $\ln\rho$ piece is the universal Coulomb phase; observable quantities
use renormalized (short-range) phase differences where the logarithm
cancels. In the flat limit ($\alpha\to1$) this reproduces the standard
Dirac--Coulomb scattering phases.

For a central potential the spin-averaged Dirac cross section can
be expressed through phase shifts. A convenient route is to separate:
\begin{itemize}
\item The pure Coulomb(flat) piece, which has a closed form (Mott),
\item Plus the topology-induced part from ($\alpha\ne1$) encoded in the
change of partial-wave phases.
\end{itemize}
For a given energy $E$, so 
\begin{equation}
k=\sqrt{E^{2}-m^{2}}\label{37}
\end{equation}
with $\beta=k/E$, the flat-space Mott differential cross section
is
\begin{equation}
\left.\frac{d\sigma}{d\Omega}\right|=\frac{\eta^{2}}{4k^{2}\sin^{4}(\tfrac{\theta}{2})}\Big(1-\beta^{2}\sin^{2}(\tfrac{\theta}{2})\Big),\label{38}
\end{equation}
which is the relativistic correction to Rutherford. To capture the
string effect we renormalize the Coulomb phases by the deficit:
\begin{equation}
\sigma_{\ell}(\eta)=\arg\Gamma(\ell+1+i\eta),\qquad\eta_{{\rm rel}}=\eta\frac{E}{k}.\label{39}
\end{equation}
A practical renormalized topological phase (vanishing at $\alpha=1$\}
is
\begin{equation}
\delta_{\ell}^{{\rm ren}}(\alpha)=\Big[\sigma_{\ell/\alpha}(\eta_{{\rm rel}})-\sigma_{\ell}(\eta_{{\rm rel}})\Big]+\frac{\pi}{2}\left(\frac{\ell+\tfrac{1}{2}}{\alpha}-\ell-\tfrac{1}{2}\right).\label{41}
\end{equation}
Using this in a short-range partial-wave sum gives a finite correction
to the amplitude,
\begin{equation}
f_{{\rm topo}}(\theta;\alpha)=\frac{1}{2ik}\sum_{\ell=0}^{\infty}(2\ell+1)\left(e^{2i\delta_{\ell}^{{\rm ren}}(\alpha)}-1\right)P_{\ell}(\cos\theta)\label{42}
\end{equation}
and the total unpolarized differential cross section is modeled by
\begin{equation}
\frac{d\sigma}{d\Omega}(\theta,E,\alpha)=\Big|f_{{\rm C}}^{{\rm flat}}(\theta;E)+f_{{\rm topo}}(\theta;E,\alpha)\Big|^{2}\Big(1-\beta^{2}\sin^{2}\tfrac{\theta}{2}\Big),\label{43}
\end{equation}
where $f_{{\rm C}}^{{\rm flat}}$)is the standard Coulomb amplitude
including the long-range phase (so that the forward $1/\sin^{2}(\theta/2$)
tail is exact). This construction reproduces Mott when $\alpha=1$
and adds the $\alpha$-dependent diffraction pattern through $\delta_{\ell}^{{\rm ren}}$. 

Because Coulomb is long-range, the total cross section diverges from
the forward region is 
\begin{equation}
\sigma(>\theta_{\min})=2\pi\int_{\theta_{\min}}^{\pi}\frac{d\sigma}{d\Omega}(\theta)\sin\theta d\theta.\label{44}
\end{equation}

\subsection{Balanced Torsion $j^{2}=0,|J_{z}|=|J_{t}|=J$}

The gamma matrices include identical torsion terms for $t$ and $z$
(Eq. (16)), inducing a coupled energy-momentum shift. The effective
$l_{eff}=l+1/2-4GJ(E+k)$.

The radial system becomes:
\begin{equation}
\frac{dF}{d\rho}+\frac{\kappa_{\text{eff}}(\alpha)}{\rho}F=(E+m-V(\rho))G,\label{45}
\end{equation}
\begin{equation}
\frac{dG}{d\rho}-\frac{\kappa_{\text{eff}}(\alpha)}{\rho}G=(E-m+V(\rho))F.\label{46}
\end{equation}
The second-order equation for $F(\rho)$ ($S=V$ case) is:
\begin{equation}
\left[-\frac{d^{2}}{d\rho^{2}}+\frac{\kappa_{\text{eff}}(\alpha)(\kappa_{\text{eff}}(\alpha)-1)}{\rho^{2}}-\frac{2\eta(E+m)}{\rho}\right]F(\rho)=(E^{2}-m^{2})F(\rho).\label{47}
\end{equation}
The solution is 
\begin{equation}
F(\rho)=N(2k\rho)^{\gamma}e^{-ik\rho}{}_{1}F_{1}\left(\gamma+i\eta(E+m)/k,2\gamma+1,2ik\rho\right),\,\label{48}
\end{equation}
where
\begin{equation}
\gamma=\sqrt{\kappa_{\text{eff}}^{2}-[\eta(E+m)/k]^{2}}\label{49}
\end{equation}
The phase shift is:
\begin{equation}
\delta_{l}=\frac{\pi}{2}(l-\kappa_{\text{eff}}(\alpha))+\arg\Gamma\left(\kappa_{\text{eff}}(\alpha)+1-i\eta(E+m)/k\right)+\eta(E+m)/k\ln(2k\rho).\label{50}
\end{equation}
For $S=-V$, replace $\kappa_{eff}\rightarrow\kappa_{eff}+1$ in the
centrifugal term and adjust $\lambda=\eta(E-m)$.

\subsection{Purely Spinning Cosmic String $J_{z}=0$}

Here, only temporal torsion contributes (Eq. (18)), yielding $l_{eff}=l+1/2-4GJ_{t}E$
(energy-dependent frame-dragging shift, independent of $k$). The
radial equations mirror Case 1, with $\kappa_{eff}(\alpha)=(l+1/2-4GJ_{t}E)/\alpha$.
The solution and phase shift follow identically, substituting this
$\kappa_{eff}$. This case introduces rotation-induced asymmetry in
scattering amplitudes, enhancing for higher $E$.

\subsection{Screw Dislocations $J_{t}=0$}

Spatial torsion dominates (Eq. (20)), shifting $l_{eff}=l+1/2-4GJ_{z}k$
(momentum-dependent, akin to a Burgers vector effect). The radial
form is as above, with $\kappa_{eff}(\alpha)=(l+1/2-4GJ_{z}k)/\alpha$.
Phase shifts are analogous, with torsion modifying low-$k$ states
more prominently.

\subsection{General Case: Torsion and Curvature-Modified Spacetime}

For arbitrary $J_{t}$ and $J_{z}$ (Eq. (23)), the combined shift
is $l_{eff}=l+1/2-4G(J_{t}E+J_{z}k)$, encapsulating both frame-dragging
and dislocation effects. The radial equation includes the radial cutoff
$\rho>\rho_{c}=(4G/\alpha)\sqrt{(J_{t})^{2}-(J_{z})^{2}}$ for $J_{t})^{2}>(J_{z})^{2}$,
imposing hard-wall boundary conditions at $\rho_{c}$. The phase shift
gains an additional geometric contribution from $\rho_{c}$, lifting
degeneracies in low-$l$ states.
\selectlanguage{english}%

\paragraph{Discussion}

\selectlanguage{american}%
The topological defect parameters ($\alpha,J_{t},J_{z}$) alter the
density of states and phase shifts, leading to modified scattering
cross sections 
\begin{equation}
\sigma=(4/k)\sum_{l}\sin^{2}\delta_{l}\label{51}
\end{equation}
For small $\alpha$, the phase shift has an additional Aharonov-Bohm-like
contribution $\delta_{AB}\approx(1-\alpha)\pi/4$ for Dirac particles.

The shifts in $l_{eff}$ introduce energy- and momentum-dependent
modifications to phase shifts, leading to altered cross sections $\sigma\propto\sum_{l}\sin^{2}\delta_{l}$.

For small topological parameters, this yields Aharonov-Bohm-like terms
\begin{equation}
\delta_{AB}\approx(1-\alpha)\pi l/2+2\pi G(J_{t}E+J_{z}k)\label{52}
\end{equation}
, observable in analog systems (e.g., graphene with defects mimicking
strings). Torsion $J_{z}$ couples to fermion chirality, potentially
distinguishable from rotation $J_{t}$ via polarized beams.

\begin{figure}
\selectlanguage{english}%
\subfloat[Differential cross section $d\sigma/d\Omega$ versus scattering angle
$\theta$ in the presence of a Coulomb potential]{\includegraphics[scale=0.35]{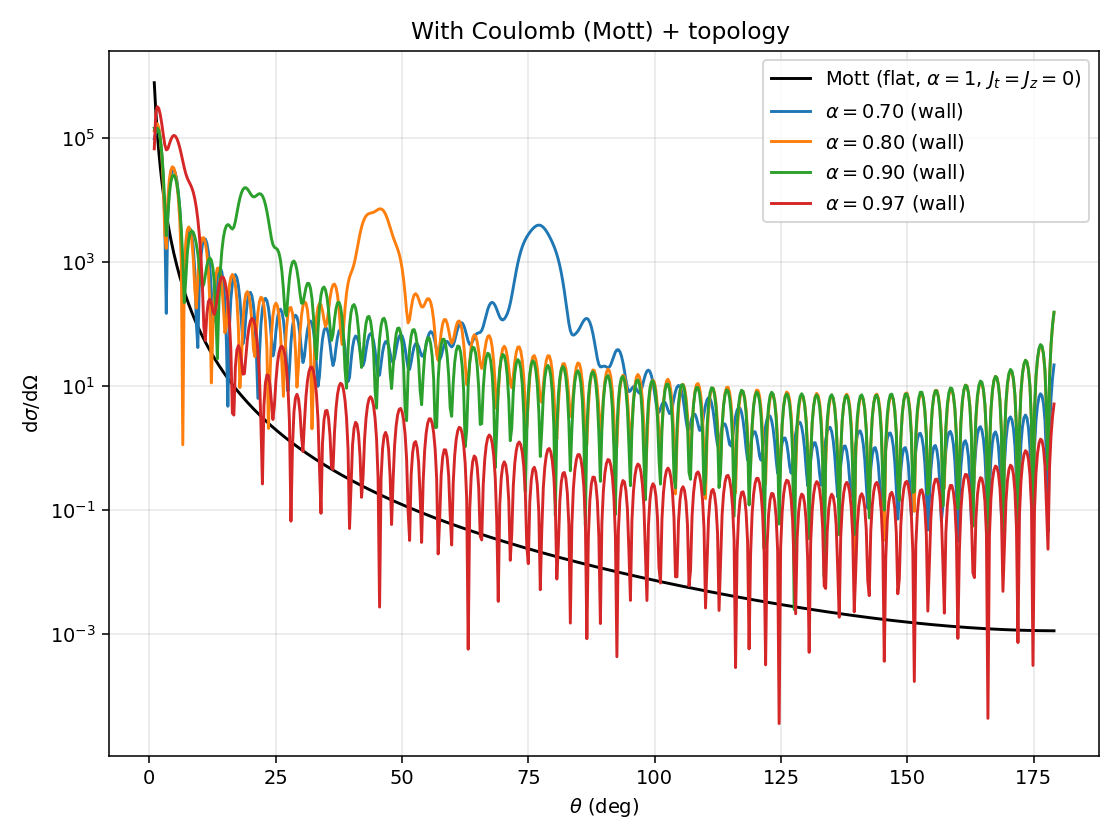}

}\subfloat[Total cross section $\sigma$ vs. energy $E/m$]{\includegraphics[scale=0.35]{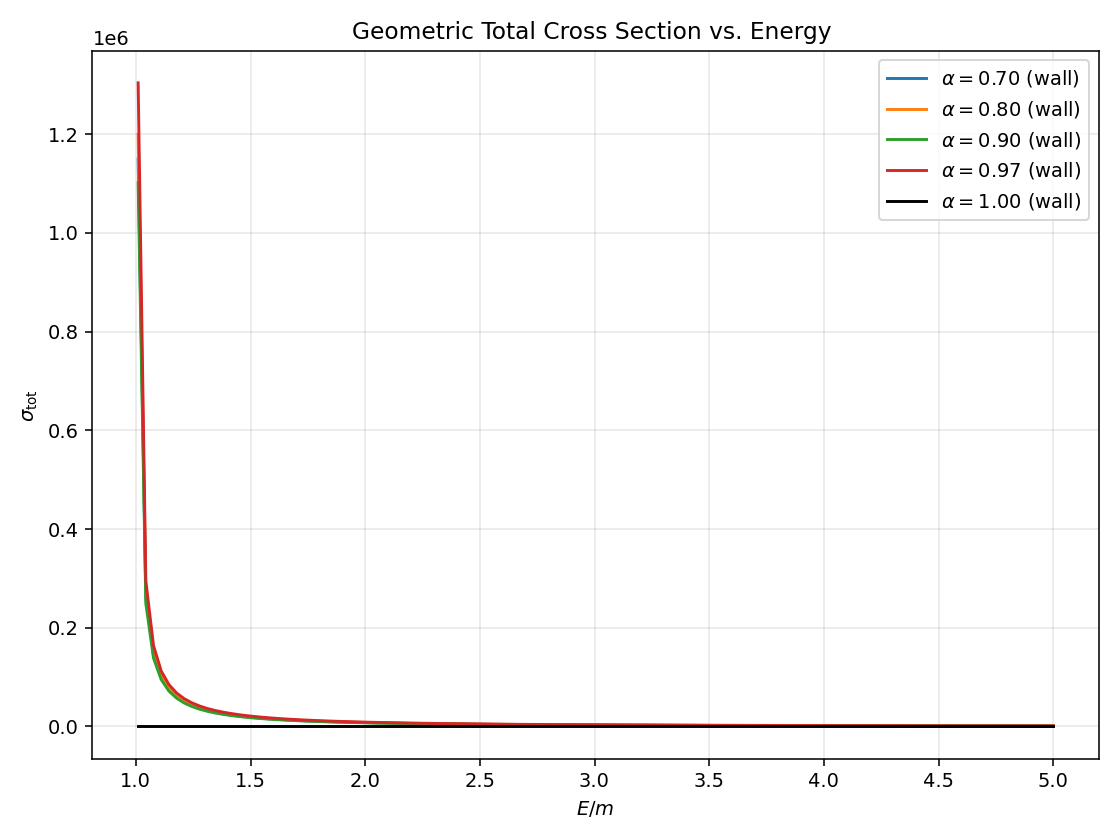}

}

\caption{Differential and Total Cross Sections for Spin-1/2 Scattering in
Pure Cosmic String Spacetime (No Torsion or Rotation)}\label{fig:1}

\selectlanguage{american}%
\end{figure}
\begin{figure}
\selectlanguage{english}%
\subfloat[Differential cross section $d\sigma/d\Omega$ versus scattering angle
$\theta$ in the presence of a Coulomb potential]{\includegraphics[scale=0.35]{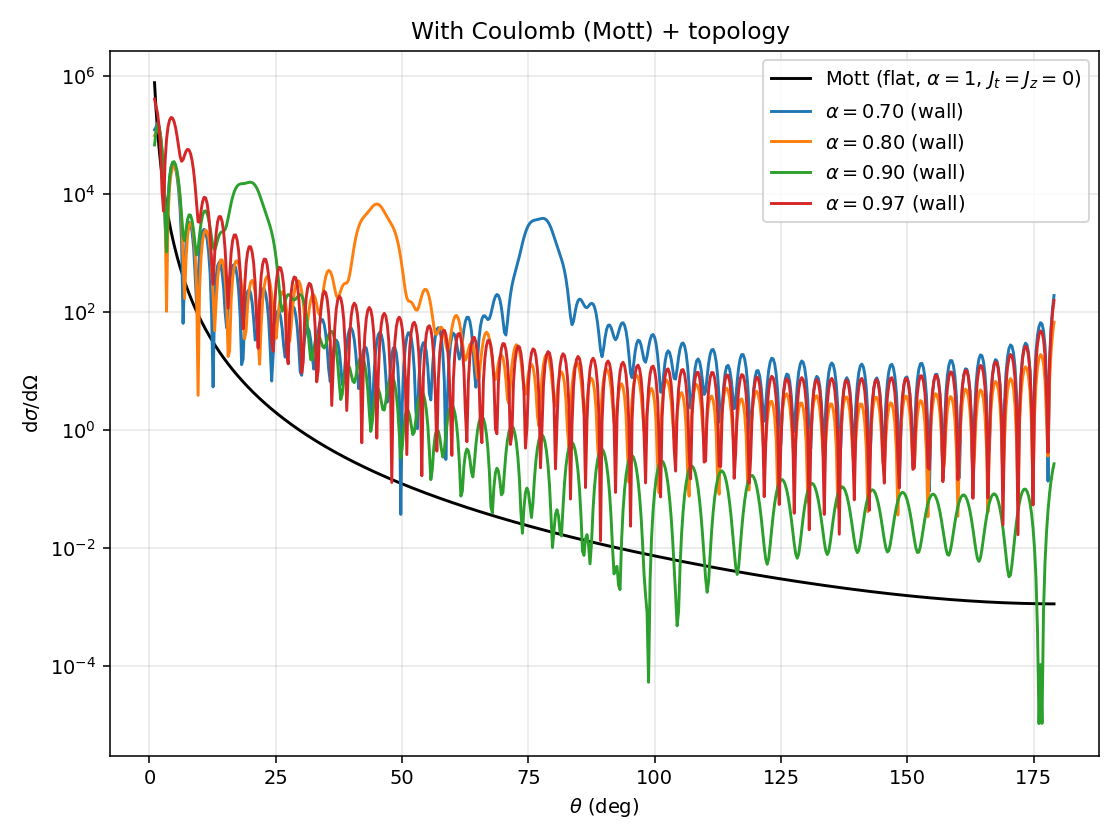}

}\subfloat[Total cross section $\sigma$ vs. energy $E/m$]{\includegraphics[scale=0.35]{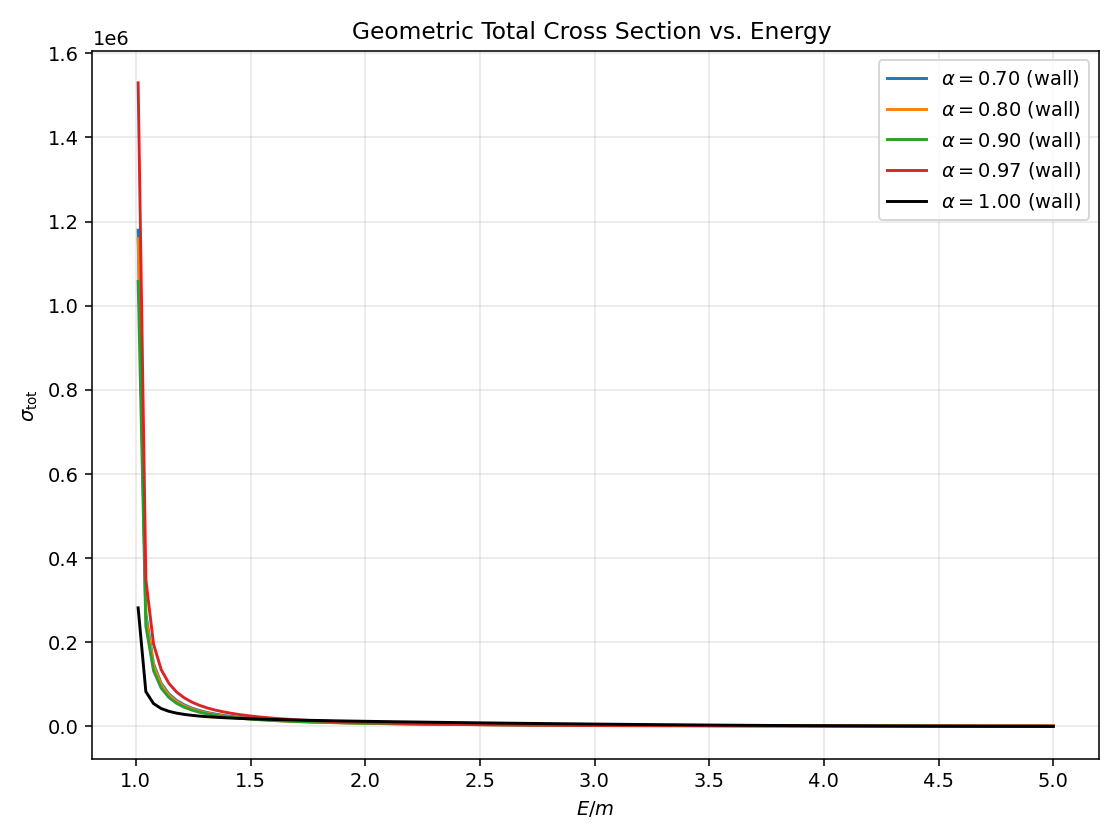}

}

\caption{Differential and Total Cross Sections for Spin-1/2 Scattering in
Balanced Torsion Cosmic String Spacetime (J\_t = J\_z)}\label{fig:2}
\selectlanguage{american}%
\end{figure}

\begin{figure}
\selectlanguage{english}%
\subfloat[Differential cross section $d\sigma/d\Omega$ versus scattering angle
$\theta$ in the presence of a Coulomb potential]{\includegraphics[scale=0.35]{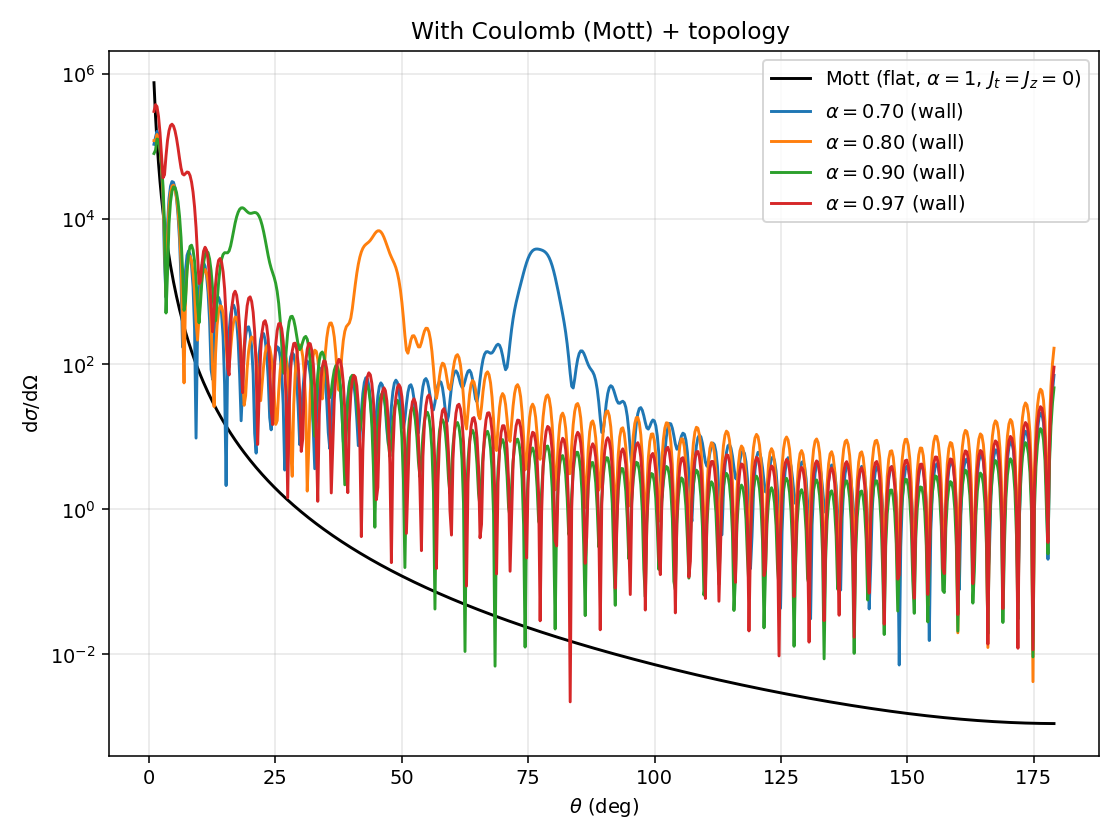}

}\subfloat[Total cross section $\sigma$ vs. energy $E/m$]{\includegraphics[scale=0.35]{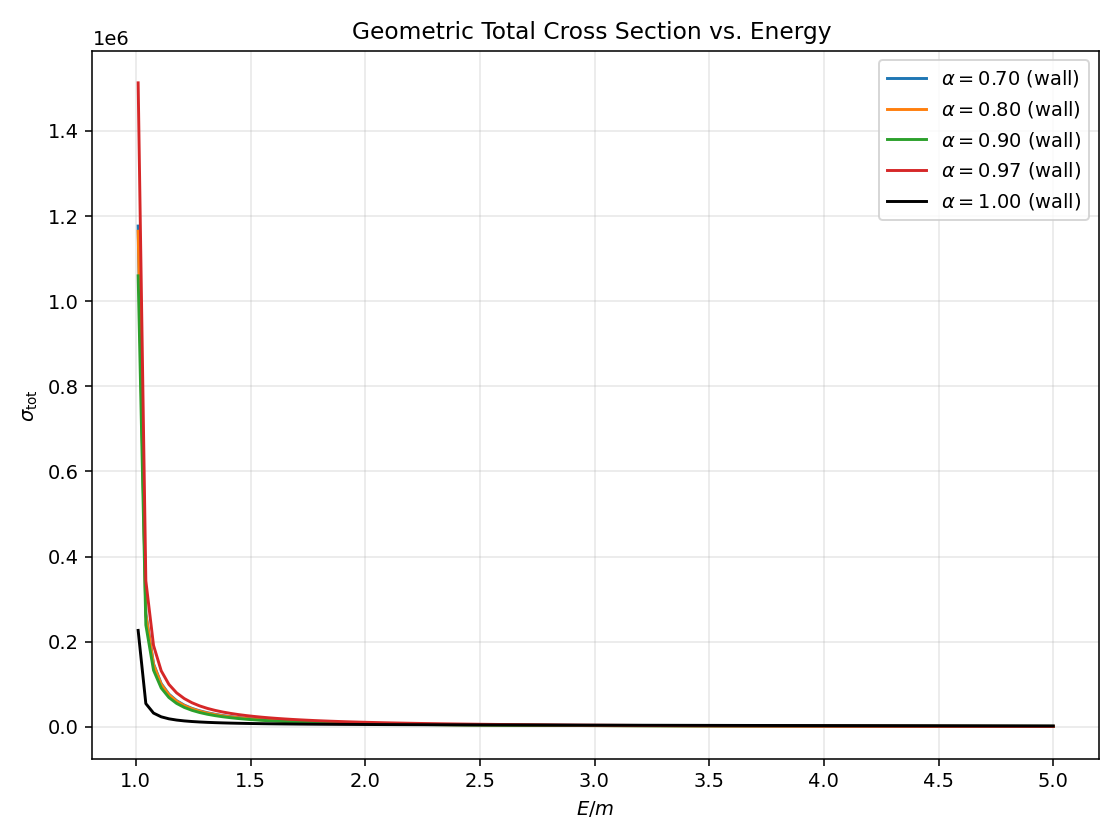}

}

\caption{Differential and Total Cross Sections for Spin-1/2 Scattering in
Pure Spinning Cosmic String Spacetime (J\_z = 0)}\label{fig:3}
\selectlanguage{american}%
\end{figure}
\begin{figure}
\selectlanguage{english}%
\subfloat[Differential cross section $d\sigma/d\Omega$ versus scattering angle
$\theta$ in the presence of a Coulomb potential]{\includegraphics[scale=0.35]{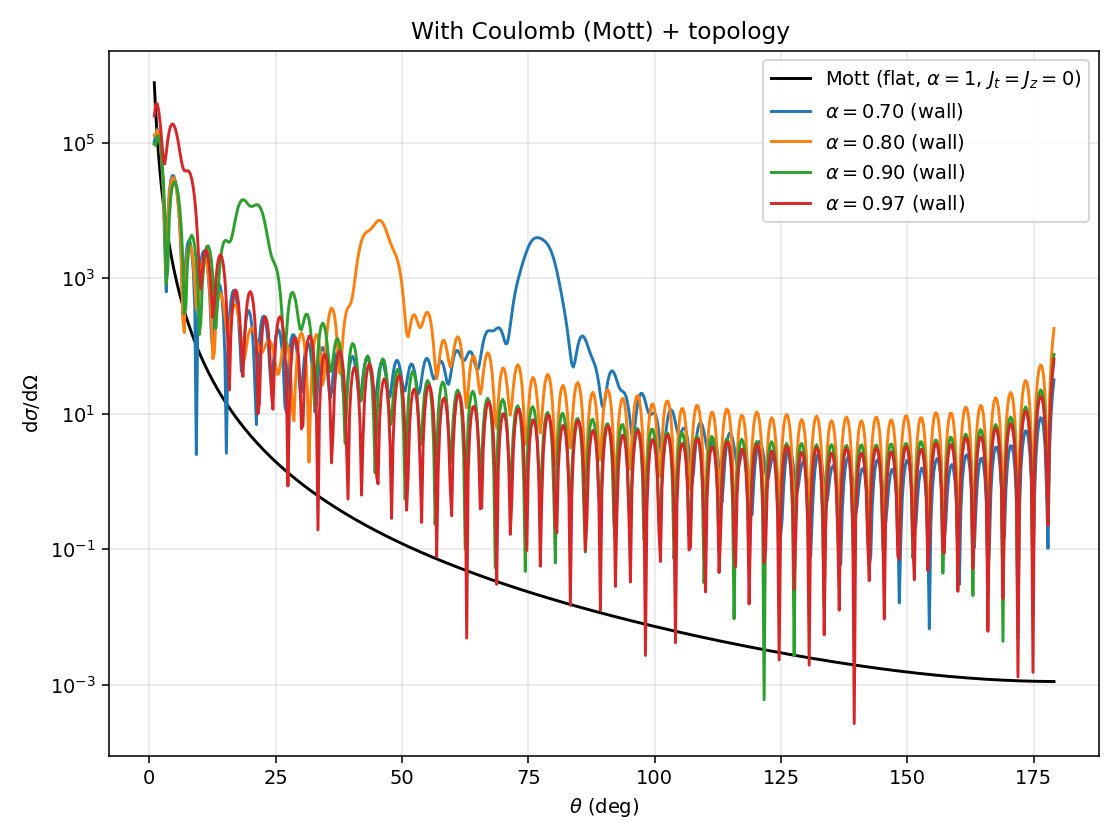}

}\subfloat[Total cross section $\sigma$ vs. energy $E/m$]{\includegraphics[scale=0.35]{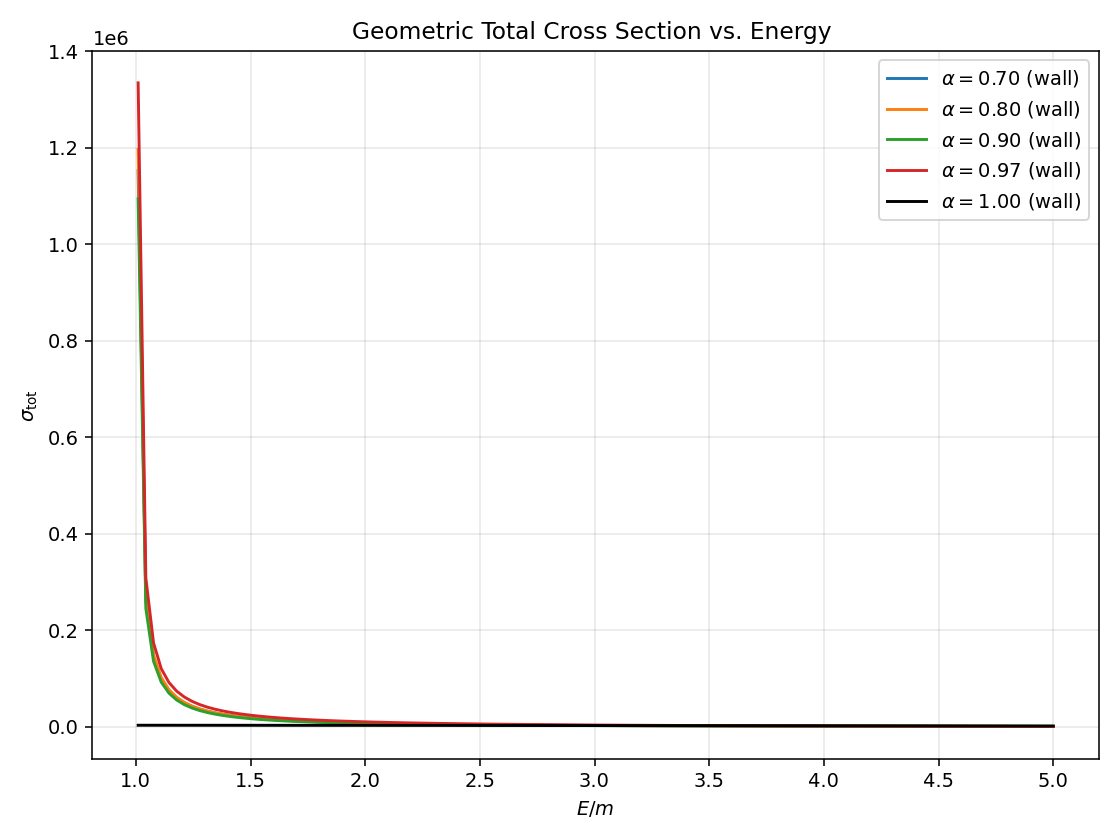}

}

\caption{Differential and Total Cross Sections for Spin-1/2 Scattering in
Screw Dislocation Cosmic String Spacetime (J\_t = 0)}\label{fig:4}
\selectlanguage{american}%
\end{figure}

\section{Geometric scattering in no electromagnetic potential}

With $A_{\mu}=0$, the Dirac equation in the stationary, axially symmetric
background reads

\begin{equation}
\Big[i\,\gamma^{\mu}(x)\big(\partial_{\mu}-\Gamma_{\mu}(x)\big)-m\Big]\Psi(x)=0,\label{53}
\end{equation}
with 
\begin{equation}
\gamma^{\mu}(x)=e^{\mu}{}_{a}(x)\gamma^{a}\label{54}
\end{equation}
 and the spin connection $\Gamma_{\mu}(x)$ built from a suitable
tetrad $e^{\mu}{}_{a}(x)$. 

We use cylindrical coordinates $(t,\rho,\theta,z)$ and the separation
ansatz
\begin{equation}
\Psi(t,\rho,\theta,z)=e^{-iEt}\,e^{il\theta}\,e^{ik_{z}z}\,\begin{pmatrix}F(\rho)\\[2pt]
G(\rho)
\end{pmatrix},\qquad l\in\mathbb{Z}.\label{eq:ansatz}
\end{equation}
Defining the transverse momentum $k_{\rho}$ by $E^{2}=m^{2}+k_{z}^{2}+k_{\rho}^{2}$,
the geometric data enter through
\begin{equation}
\ell_{\mathrm{eff}}=l+\tfrac{1}{2}-4G\big(J_{t}E+J_{z}k\big),\qquad\kappa_{\mathrm{eff}}(\alpha)=\ell_{\mathrm{eff}}/\alpha,\qquad k=\sqrt{k_{\rho}^{2}+k_{z}^{2}}.\label{eq:keff}
\end{equation}

\paragraph{Radial equations and solutions}

The coupled first-order radial system is
\begin{align}
F'(\rho)+\frac{\kappa_{\mathrm{eff}}(\alpha)}{\rho}\,F(\rho) & =(E+m)\,G(\rho),\label{eq:rad1}\\
G'(\rho)-\frac{\kappa_{\mathrm{eff}}(\alpha)}{\rho}\,G(\rho) & =(E-m)\,F(\rho).\label{eq:rad2}
\end{align}
Eliminating one component yields Bessel-type equations. Introducing
\begin{equation}
\nu_{\pm}=\Big|\kappa_{\mathrm{eff}}(\alpha)\mp\tfrac{1}{2}\Big|,\label{53-1}
\end{equation}
the regular radial solutions are
\begin{equation}
F(\rho)\propto J_{\nu_{-}}(k_{\rho}\rho),\qquad G(\rho)\propto J_{\nu_{+}}(k_{\rho}\rho).\label{54-1}
\end{equation}
For $J_{t}^{2}>J_{z}^{2}$, positivity of $g_{\theta\theta}$ imposes
the geometric cutoff
\begin{equation}
\rho>\rho_{c}=\frac{4G}{\alpha}\sqrt{J_{t}^{2}-J_{z}^{2}},\label{55}
\end{equation}
and a boundary condition at $\rho=\rho_{c}$ (Dirichlet is natural)
must be enforced; this produces a boundary-induced phase in scattering.

\paragraph{Phase shifts, amplitude, and cross sections in 2D}

Using the large-$\rho$ Hankel form, the partial-wave \$S\$-matrix
is diagonal:

\begin{equation}
S_{l}=e^{2i\delta_{l}},\qquad\delta_{l}(\alpha,J_{t},J_{z};E,k)=\frac{\pi}{2}\Big(l-\kappa_{\mathrm{eff}}(\alpha)\Big)+\delta_{{\rm wall}}(\rho_{c}),\label{eq:56}
\end{equation}
where $\delta_{{\rm wall}}$ arises only when the cutoff is active.

Expanding the incident plane wave in angular harmonics,
\begin{equation}
e^{ik_{\rho}\rho\cos\theta}=\sum_{l=-\infty}^{\infty}i^{l}J_{l}(k_{\rho}\rho)\,e^{il\theta},\label{57}
\end{equation}
and matching outgoing pieces, the 2D scattering amplitude reads
\begin{equation}
f(\theta)=\frac{1}{\sqrt{2\pi k_{\rho}}}\sum_{l=-\infty}^{\infty}\big(S_{l}-1\big)e^{il\theta}=\frac{1}{\sqrt{2\pi k_{\rho}}}\sum_{l=-\infty}^{\infty}\Big(e^{2i\delta_{l}}-1\Big)e^{il\theta},\label{eq:58}
\end{equation}
with differential cross section per unit length
\begin{equation}
\frac{d\sigma}{d\theta}=|f(\theta)|^{2},\qquad\sigma_{{\rm tot}}=\frac{4}{k_{\rho}}\,\mathrm{Im}\,f(0)=\frac{4}{k_{\rho}}\sum_{l=-\infty}^{\infty}\sin^{2}\delta_{l}.\label{eq:59}
\end{equation}
In the flat, torsionless limit $(\alpha,J_{t},J_{z})\to(1,0,0)$ one
recovers free Dirac scattering. For $\alpha\neq1$ and/or nonzero
$J_{t},J_{z}$, the phase \textbackslash eqref\{eq:phases\} contains
an Aharonov--Bohm--like geometric contribution, leading to characteristic
forward-angle enhancement; when $\rho_{c}$ is present it regularizes
the near-core behavior and lifts certain low-$l$ degeneracies..

\begin{figure}
\selectlanguage{english}%
\includegraphics[scale=0.35]{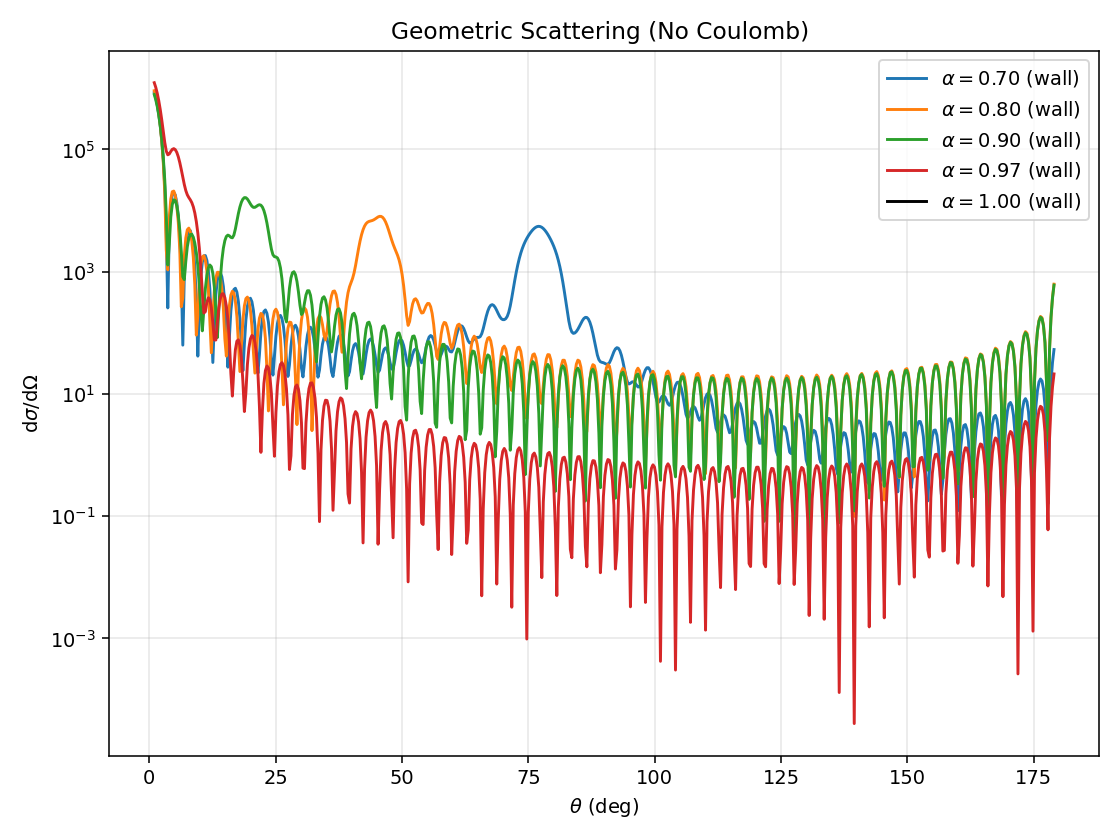} \includegraphics[scale=0.35]{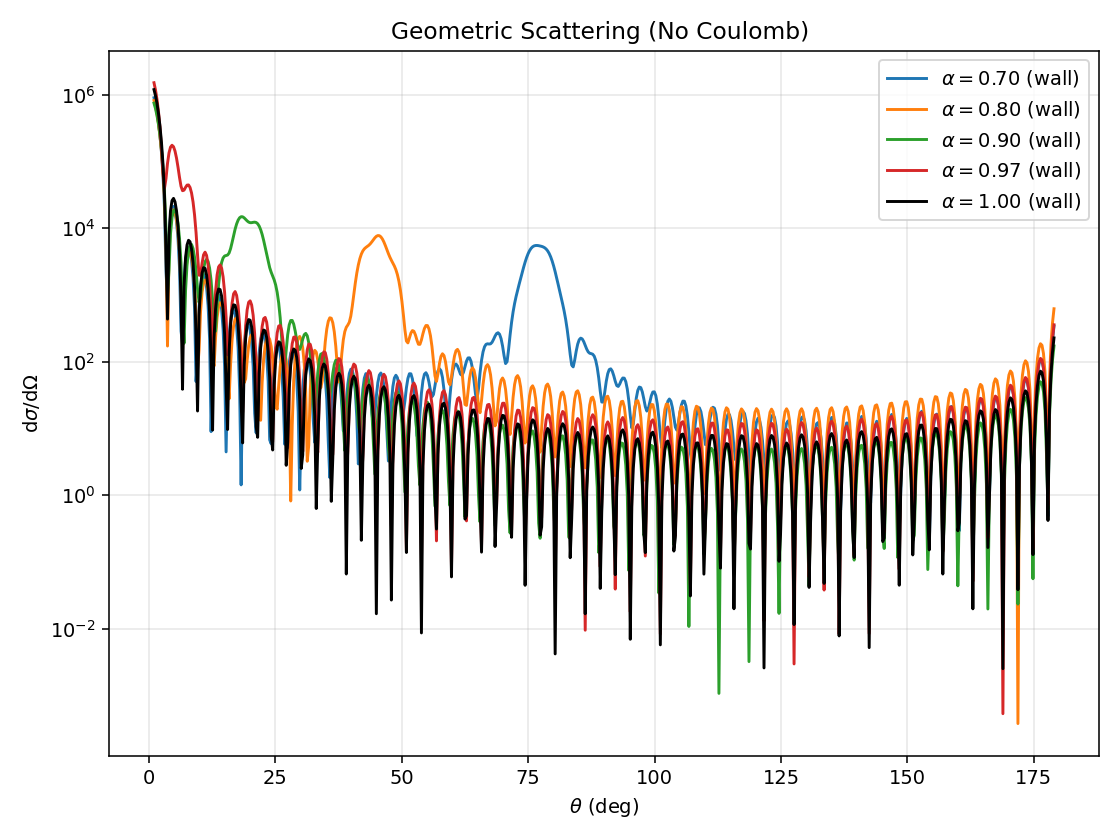}

\includegraphics[scale=0.35]{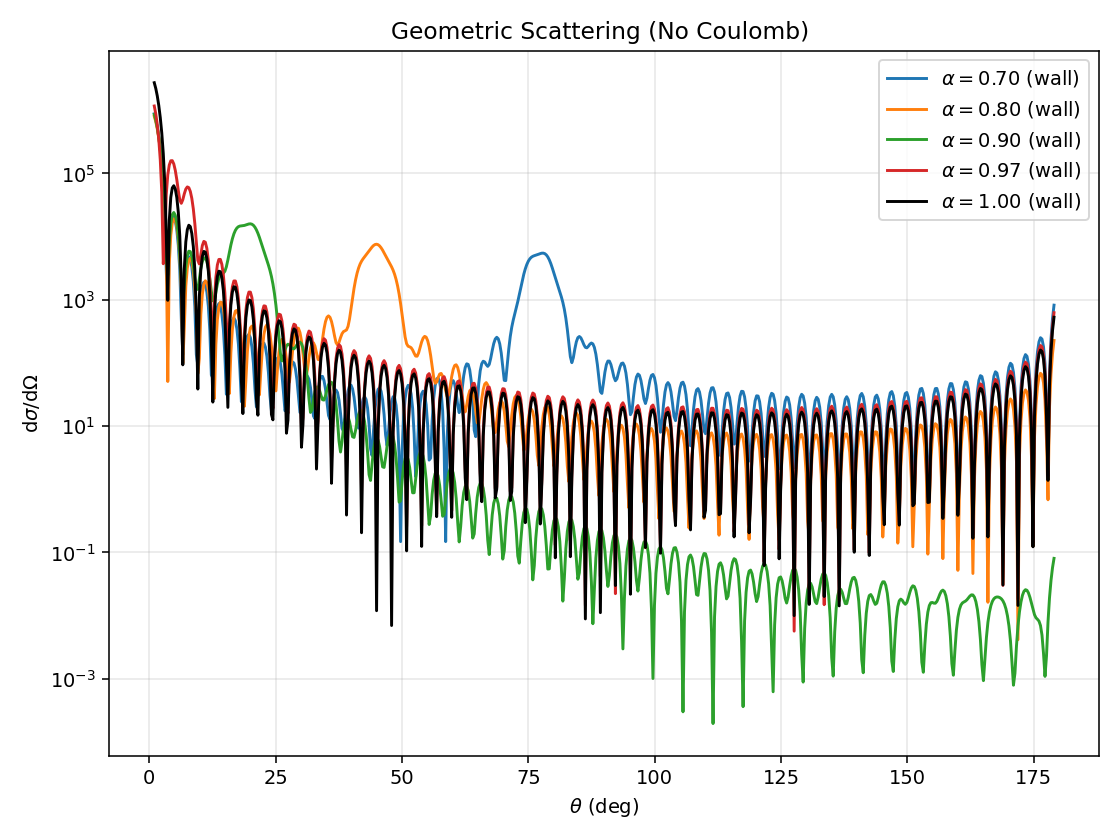}

\includegraphics[scale=0.35]{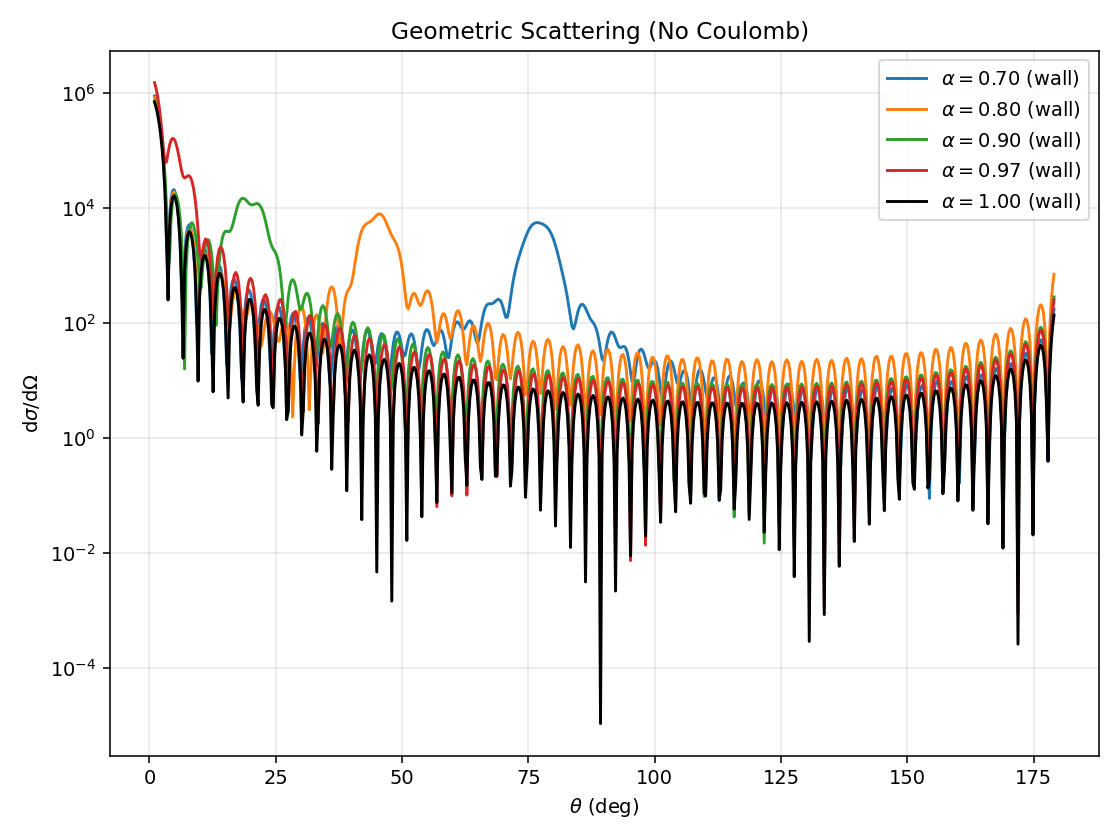}

\caption{Spacetime Geometric Effects on Scattering Cross Sections in Four
Classes of Cosmic String Models in the absence of a Coulomb potential}\label{fig:5}
\selectlanguage{american}%
\end{figure}

\section{Results and discussion}

In this section we analyze the scattering of spin-(1/2) particles
by a Coulomb potential in the spacetime of a spinning cosmic string
with spacelike disclination and dislocation, and in the Coulomb-free
limit where the scattering is purely geometric. The key quantity throughout
is the effective azimuthal quantum number, which incorporates the
conical parameter $\alpha$ and the angular momentum densities $J_{t}$
and $J_{z}$. Its dependence on the energy $E$ and longitudinal momentum
$k$ modifies the phase shifts and, consequently, the differential
and total cross sections. The figures compare several distinct geometrical
configurations to isolate the role of curvature (angular deficit),
rotation, and torsion.

\paragraph{1. Pure cosmic string background}

Figure. \ref{fig:1} illustrates the scattering in a pure cosmic string
spacetime, where only the conical defect is present and both rotation
and torsion are switched off ($J_{t}=J_{z}=0$). In this case, the
effective azimuthal quantum number reduces to a purely geometric function
of $\alpha$, and the background is static.

Panel. \ref{fig:1} (a) shows the differential cross section $d\sigma/d\Omega$
as a function of the scattering angle for different values of the
deficit parameter \textbackslash alpha. The curves all preserve the
characteristic Coulomb-dominated forward peak, reflecting the long-range
nature of the interaction and reproducing the Mott behavior in the
limit ($\alpha\to1$). However, as soon as $\alpha\neq1$, the conical
geometry induces an Aharonov--Bohm--like phase in the partial waves.
This appears as a shift of the interference maxima and minima with
respect to the flat-space result, and as a moderate enhancement or
suppression of the cross section at intermediate angles, depending
on the value of $\alpha$.

Panel. \ref{fig:1}(b) displays the corresponding (cutoff) total cross
section as a function of the particle energy. At low energies, the
deficit angle leads to a noticeable enhancement compared to the flat
case, because the geometric modification of the effective azimuthal
quantum number strongly affects the lowest partial waves, which dominate
in this regime. As the energy increases, the contribution of higher
partial waves grows and the relative impact of the conical defect
diminishes; all curves then progressively approach the flat-space
behavior. Figure 1 therefore establishes the purely conical correction
to Dirac--Coulomb scattering, which serves as a reference for the
more general geometries considered below.

\paragraph{2. Balanced torsion: spinning cosmic string with ($J_{t}=J_{z}$) }

Figure. \ref{fig:2} shows the effect of balanced torsion, where rotation
and torsion are present with equal strength, ($J_{t}=J_{z}=J$), while
the conical defect is still controlled by $\alpha$. In this case
the effective azimuthal quantum number becomes an explicit function
of both the energy ($E$ and the longitudinal momentum $k$. As a
result, the geometry is no longer a static deformation of flat space,
but couples dynamically to the kinematics of the scattered particle.

The angular distributions in Fig. \ref{fig:2}(a) exhibit much stronger
distortions than in the pure-string case. The positions of the diffraction-like
peaks and minima are significantly shifted, and their relative heights
can be either enhanced or suppressed depending on the sign and magnitude
of $J$. While the forward peak remains dominated by the Coulomb potential,
the pattern at finite angles carries clear signatures of the balanced
torsion. The comparison with Fig. \ref{fig:1}(a) shows that this
coupling to $(E,k)$ is crucial: torsion and rotation introduce a
tunable, energy-dependent geometric phase on top of the conical one.

Figure. \ref{fig:2}(b) presents the total cross section as a function
of the energy for different values of $J$. The overall trend is again
a decrease with increasing energy, but the low-energy behavior is
more sensitive than in Fig. \ref{fig:1}. The torsion-dependent term
in the effective azimuthal quantum number produces a stronger shift
of the low-$l$ phase shifts, so that the total cross section can
be either amplified or reduced compared with the pure cosmic string,
depending on the sign of $J$. This demonstrates that balanced torsion
acts as an additional “control knob” to tune the Dirac--Coulomb scattering
in such topologically non-trivial spacetimes.

\paragraph{3. Isolating rotation and dislocation effects }

Figures. \ref{fig:3} and \ref{fig:4} are devoted to disentangling
the separate contributions of rotation $J_{t}$ and screw dislocation
$J_{z}$.

In Figure. \ref{fig:3}, only the spinning character of the string
is retained: ($J_{t}\neq0$) and ($J_{z}=0$). The effective azimuthal
quantum number now depends on the energy $E$ but not directly on
the longitudinal momentum $k$. The angular distributions in Fig.
\ref{fig:3}(a) reveal an energy-dependent shift of the interference
pattern induced purely by frame dragging. Compared with Figure. \ref{fig:1},
the positions of the maxima and minima move as $E$ changes, even
for fixed $\alpha$; this is the hallmark of rotational effects. The
corresponding total cross section in Fig. \ref{fig:3}(b) shows how
this energy dependence modifies the high-energy tail: the decay of
$\sigma_{\text{tot}}(E)$ remains Coulomb-like, but its rate and normalization
are noticeably altered by rotation alone.

Figure. \ref{fig:4}, in contrast, considers a pure screw dislocation:
($J_{t}=0$) and ($J_{z}\neq0$). Here the effective azimuthal quantum
number depends on the longitudinal momentum $k$, which encodes the
influence of the Burgers-vector-like torsion. The differential cross
section in Fig. \ref{fig:4}(a) displays distortions that are most
pronounced at low momenta $k$, where the influence of the dislocation
is strongest. At higher momenta, the curves gradually approach those
of the pure string, in agreement with the expectation that short wavelengths
become less sensitive to the underlying torsional defect. The total
cross section shown in Fig. \ref{fig:4}(b) reflects this behavior:
the largest deviations from the pure-string result occur at low energies
(small (k)), while at high energies the effect of the screw dislocation
becomes comparatively small.

Taken together, Figures \ref{fig:3} and \ref{fig:4} clarify that
rotation $J_{t}$ and screw dislocation $J_{z}$ imprint qualitatively
distinct signatures on both the angular and energy dependence of the
cross section. Rotation primarily introduces an energy-driven phase
shift, while torsion associated with the dislocation leads to a momentum-driven
modification that is especially relevant at low $k$.

\paragraph{4. Geometric scattering without Coulomb interaction}

Finally, Figure 5 addresses the Coulomb-free limit, in which the scattering
is governed solely by the geometry and topology of the spinning cosmic
string spacetime. In this regime, the scattering amplitude and phase
shifts reduce to purely geometric quantities, depending only on the
effective azimuthal quantum number and the boundary conditions at
the cylindrical cutoff.

The figure compares the differential cross sections $d\sigma/d\theta$
for the four configurations considered previously: (i) pure cosmic
string, (ii) balanced torsion ($J_{t}=J_{z}$), (iii) pure spinning
string ($J_{t}\neq0,J_{z}=0$), and (iv) pure screw dislocation ($J_{t}=0,J_{z}\neq0$).
In the pure-string case, the conical defect alone produces an Aharonov--Bohm--like
angular dependence with a forward enhancement characteristic of topological
scattering. The inclusion of nonzero $J_{t}$ and/or $J_{z}$ introduces
additional shifts in the geometric phase, leading to visible changes
in the angular distributions: diffraction peaks move, their relative
intensities change, and in some cases new structures appear.

This comparison highlights that even in the absence of any long-range
potential, the nontrivial geometry of the spinning cosmic string spacetime
is sufficient to generate rich scattering patterns. The Coulomb field,
when present, superposes on these geometric effects but does not obscure
them; instead, it amplifies their phenomenological relevance by providing
a strong, experimentally familiar background against which the geometric
distortions can be measured.

Across all figures, a consistent picture emerges: the scattering of
spin-(1/2) particles in a spinning cosmic string spacetime is controlled
by an effective azimuthal quantum number that encodes curvature, rotation,
and torsion. The conical defect $\alpha$ produces Aharonov--Bohm--type
modifications of the Dirac--Coulomb cross section; rotation $J_{t}$
and screw dislocation $J_{z}$ then add energy- and momentum-dependent
contributions that can significantly reshape both angular distributions
and total cross sections. The Coulomb-free results confirm that these
are genuinely geometric and topological effects, which persist even
in the absence of conventional interactions.
\selectlanguage{english}%

\section{Potential Experimental Implementations}

The theoretical framework presented here for Dirac scattering in spinning
cosmic-string spacetimes with disclinations and dislocations has promising
analogues in condensed-matter systems, where topological defects can
mimic gravitational phenomena. Graphene, a two-dimensional Dirac material,
is an excellent platform for realising these effects experimentally,
as electrons near the Dirac points behave as massless relativistic
fermions governed by an effective Dirac equation. Meanwhile, topological
defects such as disclinations (e.g. pentagonal or heptagonal rings
in the honeycomb lattice) induce conical curvature analogous to the
angular deficit $\alpha$ in cosmic strings. This leads to modified
electronic states and Aharonov--Bohm-like phases. Strained or defective
graphene monolayers and related Dirac metals naturally exhibit disclinations
and dislocations in the honeycomb lattice that mimic conical curvature
and screw-like torsion. Non-uniform strain generates pseudo-magnetic
fields that reproduce the geometric and gauge structure of cosmic-string
spacetimes. In this sense, the parameters $\alpha$, $J_{t}$ and
$J_{z}$ map onto lattice wedge defects, local rotations and Burgers-vector-like
dislocations. Consequently, the effective azimuthal index $\kappa_{eff}(\alpha)$
and its energy/momentum dependence encode the underlying elastic and
topological fields experienced by Dirac quasiparticles. 

A concrete implementation would involve graphene on a conical or dislocated
substrate or graphene with engineered lattice defects in the presence
of a charged impurity or a gate-induced Coulomb centre. In such devices,
the topology-renormalised Mott/Rutherford patterns predicted here
would translate into modifications to the angular dependence of electronic
scattering, and consequently to the local density of states and conductance.
These could be probed using angle-resolved transport and scanning
tunnelling microscopy (STM). In moiré superlattices formed by twisted
bilayer graphene, emergent Dirac cones and strain fields further enrich
the analogy. This allows one to simulate effective curved or expanding
backgrounds, in which analogue horizons and defect structures may
give rise to phenomena similar to those observed in cosmological particle
production.

Similar ideas have been explored in the contexts of the Dirac oscillator
and defected graphene, where strain-induced pseudo-gauge fields and
torsion-like terms modify the Dirac spectrum and the scattering properties
of quasiparticles. Beyond graphene, other Dirac and Weyl materials
such as three-dimensional Dirac semimetals with disclination lines
or screw dislocations, and topological insulators with engineered
line defects offer an extended arena in which the geometric cut-off
$\rho_{c}$ and the separate roles of rotation $J_{t}$ and torsion
$J_{z}$ can be investigated: line defects and helical distortions
play the role of spinning/dislocated strings, while dopants or charged
inclusions generate long-range Coulomb fields. Additionally, since
torsion couples differently to fermion chirality than frame dragging
does, spin- or valley-polarised probes could help to disentangle the
contributions of $J_{t}$ and $J_{z}$. This would offer a realistic
way of observing the predicted geometry-induced asymmetries in a solid-state
setting, thereby testing aspects of cosmic-string scattering physics
on tabletop scales.

\section{Conclusion}

\selectlanguage{american}%
In this work, we have developed a unified description of scattering
states for spin-(1/2) particles in spinning cosmic-string spacetimes,
in the presence of a Coulomb interaction, and in the purely geometric
and spacelike disclination and dislocation limits. Starting from a
general metric characterised by the deficit parameter (\textgreek{α}),
the rotational density $J_{t}$ and the screw-dislocation parameter
$J_{z}$, we derived the curved-space Dirac equation via the tetrad
and spin-connection formalism, reducing the problem to coupled radial
equations. The geometry enters the scattering problem through an effective
azimuthal quantum number that depends on $\alpha$, $J_{t}$, $J_{z}$,
$E$, and $k$, and, in strongly rotating regimes, through a geometric
radial cutoff $\rho_{c}$ that removes noncausal regions and acts
as a hard-wall boundary. Together, these ingredients control the phase
shifts and hence the observable cross sections.

For the Coulomb problem, we obtained analytical scattering solutions
in several distinct backgrounds: a pure cosmic string; a spinning
string with balanced torsion $J_{t}=J_{z}$; a purely spinning string
$J_{z}=0$; a screw-dislocation string $J_{t}=0$; and the general
case. In all of these cases, the long-range Coulomb force provides
the familiar Mott/Rutherford backbone, while the defect parameters
impart additional Aharonov--Bohm--like phases. Our results demonstrate
that the conical defect primarily generates a topology-induced modulation
of the angular interference pattern and a low-energy enhancement of
the total cross section. In contrast, rotation and torsion introduce
genuine shifts in the effective azimuthal index that depend on energy
and momentum. In particular, frame dragging associated with $J_{t}$
leads to an energy-driven distortion of the angular distribution,
while the screw dislocation associated with $J_{z}$ produces momentum-driven
modifications that are most pronounced at low $k$. When the cutoff
$\rho_{c}$ is active, an additional boundary phase emerges, eliminating
low-angular-momentum degeneracies and stabilising the near-core region.

We have also analysed the Coulomb-free regime, where scattering is
governed solely by spacetime geometry. In this limit, the amplitude
and phase shifts reduce to purely geometric quantities; however, the
resulting patterns remain highly non-trivial: the conical defect alone
yields a forward enhancement and diffraction structure that is closely
analogous to topological Aharonov--Bohm scattering. Meanwhile, non-zero
$J_{t}$ and $J_{z}$ further shift and reshape the diffraction peaks.
Comparing the Coulomb and non-Coulomb cases shows that the electromagnetic
interaction does not obscure the geometric effects, but rather amplifies
their phenomenological visibility by providing a strong, well-understood
reference against which geometry-induced deviations can be identified.

Finally, we argue that these results are relevant not only to hypothetical
cosmic strings in the early Universe, but also to condensed-matter
analogues in Dirac materials. Strained or defective graphene, as well
as three-dimensional Dirac and Weyl semimetals with line defects,
can mimic the conical curvature, rotation and torsion of spinning
or dislocated strings. In such systems, the effective azimuthal index
and geometric cutoff translate into elastic and topological fields
acting on quasiparticles. The topology-renormalised Mott/Rutherford
patterns derived here correspond to measurable changes in angular-dependent
electronic scattering, local density of states and transport. Future
work may extend our framework to include additional interactions,
such as magnetic fields or non-central potentials, investigate polarisation-resolved
and time-dependent scattering and perform numerical simulations in
fully dynamical string backgrounds. Such studies would further clarify
the interplay between geometry, topology, and quantum scattering in
astrophysical scenarios and tabletop analogue experiments.

\selectlanguage{english}%
\bibliographystyle{ChemEurJ}
\bibliography{referenceboumali_clean}

\end{document}